\documentclass[twocolumn,showpacs,preprintnumbers,amsmath,amssymb]{revtex4}
\usepackage{graphicx}
\usepackage{dcolumn}
\usepackage{xcolor}	
\usepackage{mathtools}
\usepackage{tikz}
 \usepackage{printlen}

\newcommand{\op}[1]{\ensuremath{\hat{#1}}}
\newcommand{\ladderdown}{\ensuremath{\op{a}^{\vphantom{\dagger}}}}
\newcommand{\ladderup}{\ensuremath{\op{a}^\dagger}}
\newcommand{\annihilop}{\ladderdown}
\newcommand{\creationop}{\ladderup}

\begin{document}
\bibliographystyle{apsrev}


\title{Ab initio thermodynamic results for the degenerate electron gas at finite temperature}

\author{T.~Schoof}
\author{S.~Groth}
\author{J.~Vorberger}
\author{M.~Bonitz}
\affiliation{Institut f\"ur Theoretische Physik und Astrophysik, Christian-Albrechts-Universit\"{a}t zu Kiel, D-24098 Kiel, Germany}


\date{\today}
\begin{abstract}
The uniform electron gas (UEG) at finite temperature is of key relevance for many applications in dense plasmas, warm dense matter, laser excited solids and much more. Accurate thermodynamic data for the UEG are an essential ingredient for many-body theories, in particular, density functional theory. Recently, first-principle restricted path integral Monte Carlo results became available which, however, due to the fermion sign problem, had to be restricted to moderate degeneracy, i.e. low to moderate densities with $r_s={\bar r}/a_B \gtrsim 1$. 
Here we present novel first-principle configuration PIMC results for 
electrons
for $r_s \leq 1$. We also present quantum statistical data within the $e^4$-approximation that are in good agreement with the simulations at small to moderate $r_s$.
\end{abstract}

\maketitle

Thermodynamic properties of quantum degenerate electrons are vital for the description of matter at high densities, e.g.\ \cite{continuum14,Kraus:2013,Regan:2012}, such as  dense plasmas in compact stars or planet cores, e.g.\ \cite{knudson_12, Militzer:2008, Nettelmann:2012}, as well as in laser fusion experiments at NIF, e.g.\ \cite{lindl_04, hu_11, hurricane_nif14}, 
Rochester \cite{theobald_15} or Sandia~\cite{hanson_14,hanson_14_2}. Besides, the electron component is of crucial importance for understanding the properties of atoms, molecules and existing and novel materials. 
%
The most successful approach has been density functional theory (DFT)--combined with an approximation for the exchange-correlation potential.
Its success is based on the availability of accurate {\em zero temperature} data for the UEG which is obtained from analytically known limiting cases combined with first-principle quantum Monte Carlo data~\cite{ceperley_alder}.

\begin{figure}
 \includegraphics[width=85mm]{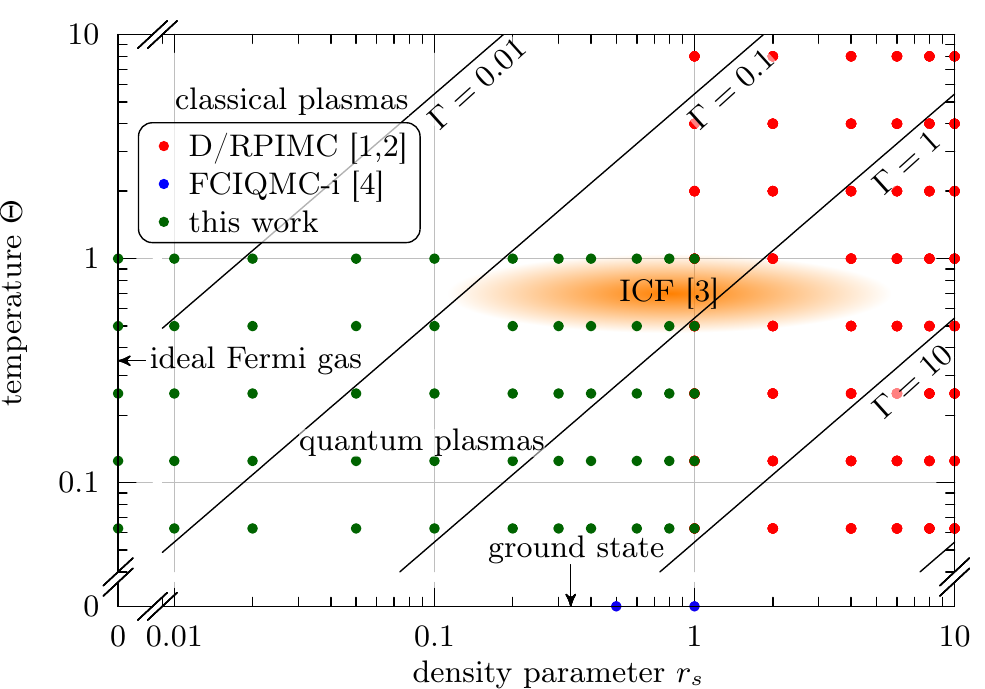}
 \caption{Density-temperature plain in the warm dense matter range. ICF: typical inertial confinement fusion parameters \cite{hu_11}. Quantum (classical) behavior dominates below (above) the line $\Theta=1$. $\Gamma=e^2/{\bar r}k_BT$  is the classical coupling parameter. 
 Red dots: available finite temperature RPIMC \cite{brown_prl13} and DPIMC \cite{filinov_14} data for the UEG. Blue dots: ground state data of Ref.~\cite{alawi_prb12}. Green dots: CPIMC and analytical results of this work.
}
 \label{fig:n-t-sketch}
\end{figure}

In recent years more and more applications have emerged where the electrons are highly excited, e.g.\ by compression of the material or by electromagnetic radiation (see above), which require to go beyond zero temperature DFT. This has led to an urgent need for accurate thermodynamic data of the UEG at {\em finite temperature}. One known limiting case is the highly degenerate ideal Fermi gas (IFG), and perturbation theory results around the IFG, starting with the Hartree-Fock and first order correlation corrections (Montroll-Ward) \cite{dewitt61,kraeft79}, are long known; for an analytical fit at high densities, see Ref.~\cite{perrot84} and for further improved approximations, such as the $e^4$-approximation, see Refs.~\cite{green-book, kremp_springer,vorberger_pre04}. These approximations break down when the Coulomb interaction energy among the electrons becomes comparable to their kinetic energy, requiring computer simulations such as path integral Monte Carlo (PIMC), e.g.\ \cite{ceperley95rmp}.
While restricted PIMC (RPIMC) results for dense multi-component quantum plasmas, e.g.\ \cite{mil-pol, militzer_06} as well as direct fermionic PIMC (DPIMC) results~\cite{filinov-etal.00jetpl,FiBoEbFo01,afilinov-etal.04pre,Bonitz_PRL05} have been available already for 15 years,
only recently finite temperature RPIMC results for the UEG (or jellium) have been obtained~\cite{brown_prl13}. It is well known that fermionic PIMC simulation in continuous space suffer from the fermion sign problem (FSP) which is known to be NP hard~\cite{troyer}. This means, with increasing quantum degeneracy, i.e.\ increasing parameter $\chi = n \lambda^3_{DB}$, which is the product of density and thermal DeBroglie wave length, $\lambda_{DB}^2=h^2[2\pi m k_BT]^{-1}$, the simulations suffer an exponential loss of accuracy. 
RPIMC formally avoids the FSP by an additional assumption on the nodes of the density matrix, however, 
it also cannot access high densities~\cite{filinov_rpimc}, $r_s < 1$ [$r_s={\bar r}/a_B$, where ${\bar r}$ is the mean interparticle distance, $n^{-1}= 4\pi {\bar r}^3/3$ and $a_B$ the Bohr radius]. Also, the quality of the simulations 
around $r_s = 1$, at low temperatures $\Theta= k_BT/E_F \le 0.125$ [$E_F$ is the Fermi energy] is unknown. However, this leaves out the high-density range that is of high importance, e.g.\ for deuterium-tritium implosions at NIF where mass densities of $400 $ gcm$^{-3}$ (up to $1596$ gcm$^{-3}$) have recently been reported~\cite{hurricane_nif14} (are expected along the implosion path~\cite{hu_11}), corresponding to $r_s\approx 0.24$ ($r_s = 0.15$), see Fig.~\ref{fig:n-t-sketch}.

The authors of Ref.~\cite{brown_prl13} also performed DPIMC simulations which confirmed that, for $\Theta < 0.5$ and $r_s \lesssim 4$, these simulations are practically not possible, see Fig.~\ref{fig:n-t-sketch}. We also mention independent recent DPIMC simulations~\cite{filinov_14} that are overall in good agreement with the data of Ref.~\cite{brown_prl13} but indicate large deviations for the lowest temperatures and $r_s \lesssim 2$ (lower energies). Finally, a recent attempt to avoid the FSP by an approximate treatment of exchange cycles~\cite{dubois14} also yielded lower energies for $r_s=1$ and $\Theta=0.125$ than the RPIMC data of Ref.~\cite{brown_prl13}.
To bridge the gap between the known analytical result for the ideal Fermi gas at $r_s=0$ and previous simulations ($r_s \gtrsim 1$) and to provide comprehensive input data for finite temperature DFT, several fits have been proposed recently~\cite{brown_prb13, karasiev_prl14}. However, they crucially depend on the quality of the underlying simulation data.

It is the purpose of this Letter to improve this situation. We present the first ab initio simulation results that avoid a simplified treatment of fermionic exchange for $r_s \lesssim 1$ and finite temperatures, $\Theta \lesssim 1.0$.
We apply the recently developed fermionic configuration path integral Monte Carlo (CPIMC) approach to the UEG and demonstrate its capabilities for $33$ spin polarized electrons in a cubic box of side length $L$ (as was studied in~\cite{brown_prb13, dubois14}). Our simulations have no sign problem for $0 \le r_s \le 0.4$ and are accurate up to $r_s=1$.

{\bf CPIMC for the UEG}. The jellium Hamiltonian in second quantization with respect to plane waves 
$\langle \vec{r}\;|\vec{k}\rangle = \frac{1}{L^{3/2}} e^{i\vec{k} \cdot \vec{r}}$ with $\vec{k}=\frac{2\pi}{L}\vec{m}$, $\vec{m}\in \mathbb{Z}^3$ 
has the familiar form (we use Rydberg units throughout) 
\begin{align}\label{eq:h} 
& \op{H} =
\sum_{i}\vec{k}_i^2 \creationop_{\vec{k}_i}\annihilop_{\vec{k}_i} + 2\smashoperator{\sum_{\substack{i<j,k<l \\ i\neq k,j\neq l}}} 
w^-_{ijkl}\creationop_{i}\creationop_{j} \annihilop_{l} \annihilop_{k} + E_M,
\\
& w^-_{ijkl} =w_{ijkl}-w_{ijlk}, \quad w_{ijkl}=\frac{4\pi e^2}{L^3}\frac{\delta_{\vec{k}_i+\vec{k}_j, \vec{k}_k + \vec{k}_l} }{\vec{k}_{ik}^2},
\label{eq:w_minus}
\end{align}
where the first (second) term describes the kinetic (interaction) energy, and $\vec{k}_{ik}=\vec{k}_{i}-\vec{k}_{k}$. The Madelung energy $E_M$ accounts for the self-interaction of the Ewald summation in periodic boundary conditions~\cite{fraser_madelung} for which we found $E_M\approx-2.837297\cdot(3/4\pi)^{\frac{1}{3}}N^{\frac{2}{3}}r_s^{-1}$. 
The operator 
$\creationop_{i}$  ($ \annihilop_{i}$) 
creates (annihilates) a particle in the orbital $|\vec{k}_i\rangle$.
In the interaction term, the $\vec{k}_i=\vec{k}_k$ and $\vec{k}_j=\vec{k}_l$ components cancel with the interactions with the positive background. 
While the complete (infinite) set of plane waves $\langle\vec{r}\;|\vec{k}_i\rangle$ forms a basis in the single-particle Hilbert space, for simulations it has to be truncated at a number $N_B$ of orbitals.

In conventional RPIMC and DPIMC, the system~(\ref{eq:h}) is treated in the coordinate representation allowing for a numerically exact description in the classical strongly coupled limit and for weak degeneracy. CPIMC~\cite{schoof_cpp_11}, in contrast, is constructed in a way that it allows for exact simulations in the opposite limit of the ideal Fermi gas, $r_s=0$ \cite{relativistic}, and at weak to moderate coupling and strong degeneracy. This is achieved by representing the $N$-electron state in second quantization~\cite{ctmc-comment} as $|\{n\}\rangle=|n_1, n_2, \dots\rangle$, where the $n_i$ are fermionic occupation numbers ($n_i=0, 1$) of the orbitals $|\vec{k}_i\rangle$.
This way, fermionic anti-symmetry is ``built in'' exactly. The partition function $Z$ and quantum-statistical expectation values, such as the internal energy $U$, are straightforwardly computed in Fock space as 
\begin{align}
Z(\Theta, r_s; N) &=  \mbox{Tr}_{|\{n\}\rangle} \, e^{-\beta {\hat H}}, 
\label{eq:z}
\\
U(\Theta, r_s; N) = \langle {\hat H} \rangle &= Z^{-1}\,\mbox{Tr}_{|\{n\}\rangle} {\hat H}\, e^{-\beta {\hat H}} ,
\label{eq:mean-h} 
\end{align}
avoiding the sum over $N!$ permutations over the canonical density operator $e^{-\beta {\hat H}}$ that is the origin of the notorious FSP in DPIMC. 
The trace is evaluated using the concept of continuous time PIMC which has been successfully applied to bosonic lattice models~\cite{prokofiev96,prokofiev98,houcke06,rombouts06}. We have generalized this concept to continuous fermionic systems with long range interactions~\cite{schoof_cpp_11,simon_springer14}.  
The main idea is to split the Hamiltonian into a diagonal, $\op{D}$, and an off-diagonal part, $\op{Y}$ and summing up the entire perturbation series of the density operator $e^{-\beta {\hat H}}$ in terms of $\op{Y}$. The final result, for the case of the UEG, is~\cite{schoof_cpp14}:
\begin{align}
Z = &
\sum_{K=0,\atop K \neq 1}^{\infty} \sum_{\{n\}}
\sum_{s_1\ldots s_{K-1}}\,\label{eq:z-kinks}
\int\limits_{0}^{\beta} d\tau_1 \int\limits_{\tau_1}^{\beta} d\tau_2 \ldots \int\limits_{\tau_{K-1}}^\beta d\tau_K 
\\\nonumber
& (-1)^K  
e^{-\sum\limits_{i=0}^{K} D_{\{n^{(i)}\}} \left(\tau_{i+1}-\tau_i\right) } 
\prod_{i=1}^{K}(-1)^{\alpha^{\phantom{-}}_{s_i}}\,w^-_{s_i}\;,
%
\\
\label{eq:d}
 D_{\{n^{(i)}\}} &= \sum_l \vec{k}_l^2 n^{(i)}_{l} + \sum_{l<k}w^-_{lklk}n^{(i)}_{l}n^{(i)}_{k} \;,
\\
\alpha^{\phantom{-}}_{s_i} &=\alpha^{(i)}_{pqrs}=\sum_{l=p}^{q-1}n^{(i-1)}_{l}+\sum_{l=r}^{s-1}n^{(i)}_{l}\;.
\nonumber
\end{align}
where $s_i=(p,q,r,s)$, with $p<q,\; r<s$ denotes a quadruple of pairwise different orbital indices.
\begin{figure}
 \begin{tikzpicture}[xscale=0.74, yscale=0.4]
\newcommand{\xrange}{10}
\newcommand{\yrange}{1}
\newcommand{\taueins}{\xrange*0.3}
\newcommand{\tauzwei}{\xrange*0.45}
\newcommand{\taudrei}{\xrange*0.8}
\newcommand{\zero}{\yrange*1}
\newcommand{\eins}{\yrange*2}
\newcommand{\zwei}{\yrange*3}
\newcommand{\drei}{\yrange*4}
\newcommand{\vier}{\yrange*5}
\newcommand{\fuenf}{\yrange*6}

\draw[->] (0,0) -- +(\xrange+0.5*\xrange/20,0) coordinate (xlabel);
\draw[->] (0,0) -- +(0,\yrange*7) coordinate (ylabel);
\foreach \i in {0,...,5} {
	\draw (-0.1,\i*\yrange+\yrange) node[left] {$\i$} -- (0.1,\i*\yrange+\yrange);
}
\foreach \i/\l in {\taueins/$\tau_1$,\tauzwei/$\tau_2$,\taudrei/$\tau_3$} {
\draw (\i,0.1) -- (\i,-0.1) node[below] {\l};
}
\draw (0,0.1) -- (0,-0.1) node[below] {$0$};
\draw (\xrange,0.1) -- (\xrange,-0.1) node[below] {$\beta$};
\node at (0.5*\xrange,-1.5) {imaginary time $\tau$};
\node[rotate=90] at (-0.8,3.5*\yrange) {orbital $i$};

\foreach \i in {1,...,6} {
\draw[semithick,dotted] (0,\i*\yrange) -- (\xrange,\i*\yrange);
}

\draw[dashed] (\tauzwei+2.0,0.5) rectangle (\tauzwei+2.75,\yrange*6.5);
\node at (\tauzwei + 3.0,\fuenf+\yrange+0.2) {$|\{n^{(2)}\}\rangle=|001110\ldots\rangle$};

\begin{scope}[thick]
\draw (0,\zero) -| (\taueins,\zwei) -| (\taudrei,\zero) -- (\xrange,\zero);
\draw (0,\eins) -| (\tauzwei,\vier) -| (\taudrei,\eins) -- (\xrange,\eins);
\draw (0,\drei) -| (\taueins,\fuenf) -| (\tauzwei,\drei) -- (\xrange,\drei);
\draw (\taueins, \drei) -- (\taueins, \zwei);
 \end{scope}
 
 \node at (\taueins, \fuenf+\yrange+0.2) { $s_1 =(2,5,0,3)$};

\end{tikzpicture}
 \caption{Typical closed path in Slater determinant (Fock) space. The state with three occupied orbitals $|{\vec k}_0\rangle, |{\vec k}_1\rangle, |{\vec k}_3\rangle$ undergoes a two-particle excitation $(s_1, \tau_1)$ that replaces the occupied orbitals $|{\vec k}_0\rangle, |{\vec k}_3\rangle$ by 
$|{\vec k}_2\rangle, |{\vec k}_5\rangle$. Two further excitations occur at $\tau_2$ and $\tau_3$.
The states at the ``imaginary times'' $\tau = 0$ and $\tau = \beta$ coincide. All possible paths contribute to the partition function $Z$, Eq.~(\ref{eq:z-kinks}).}
 \label{fig:sketch}
\end{figure}
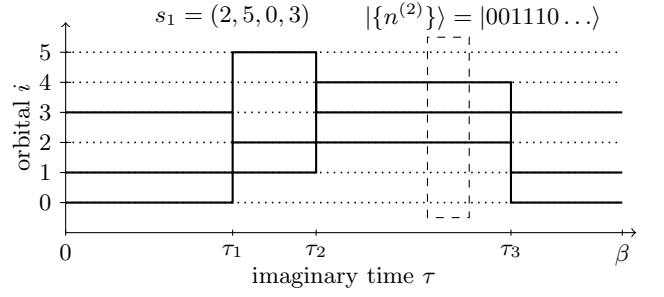

Thus the partition function is represented as a sum over 
$\beta$-periodic ``paths'' in Fock space, in imaginary time, which we illustrate in Fig.~\ref{fig:sketch}: For an ideal Fermi system a path is characterized by a single $N$-particle Slater determinant $|\{n\}\rangle$.   
For a correlated Fermi system the original determinant $|\{n\}\rangle=|\{n^{(0)}\}\rangle$ (straight horizontal lines in Fig.~\ref{fig:sketch}) is interrupted by excitations of the type
$(s,\tau)$: at time $\tau$, a pair of occupied orbitals $|{\vec k}_r\rangle, |{\vec k}_s\rangle$ 
is replaced by the previously empty pair  $|{\vec k}_p\rangle, |{\vec k}_q\rangle$. Paths differ by the number  $K$ of excitations (``kinks''), their times $\tau_1\dots \tau_k$ on the $\tau$-interval $[0,\beta]$ 
and the involved quadruples of orbitals $s_1\dots s_K$. The partition function clearly reflects this summation over the different types 
of kinks, integration over the kink times and summation over $K$ [cf.\ first line of Eq.\ (\ref{eq:z-kinks})]. 
The weight of each path [terms in the second line of Eq.~(\ref{eq:z-kinks})] is determined by the Fock state matrix elements of the Hamiltonian, where diagonal elements $D_{\{n^{(i)}\}}$, Eq.~(\ref{eq:d}), arise from the kinetic energy and the mean-field part of the Coulomb interaction, whereas off-diagonal 
elements, $(-1)^{\alpha^{\phantom{-}}_{s_i}}w^-_{s_i}$, are due to the remaining Coulomb interaction (correlation part) \cite{kink_note}.
Expression~(\ref{eq:z-kinks}) is exact for $N_B \to \infty$ allowing for ab initio thermodynamic simulations of the UEG.

Formula~(\ref{eq:z-kinks}) and similar expressions for thermodynamic observables, such as the internal energy~\cite{schoof_cpp14}, can be efficiently evaluated using Metropolis Monte Carlo. To this end, we developed an algorithm that generates all possible paths in Slater determinant space thereby assuring ergodicity. For the UEG a total of 6 different steps are required, including addition and removal of a single kink and pairs of kinks, modification of an existing kink and excitation of single orbitals. The complete set of steps and the associated MC algorithm cannot be presented here, for details see Ref.~\cite{schoof_cpp14}.

\begin{figure}
 \includegraphics{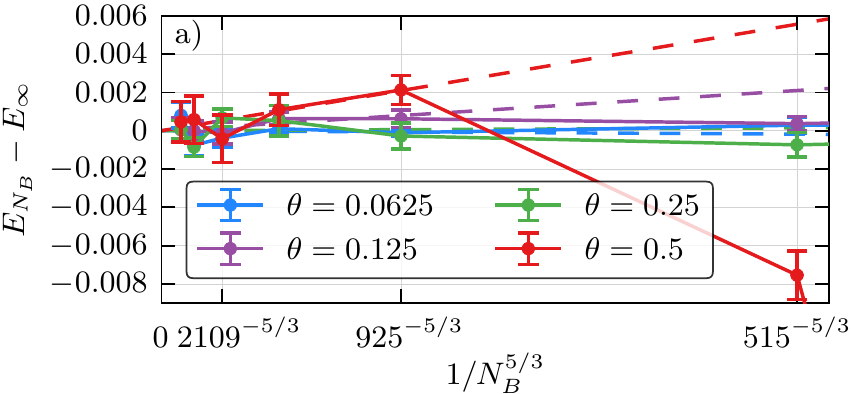}
  \includegraphics{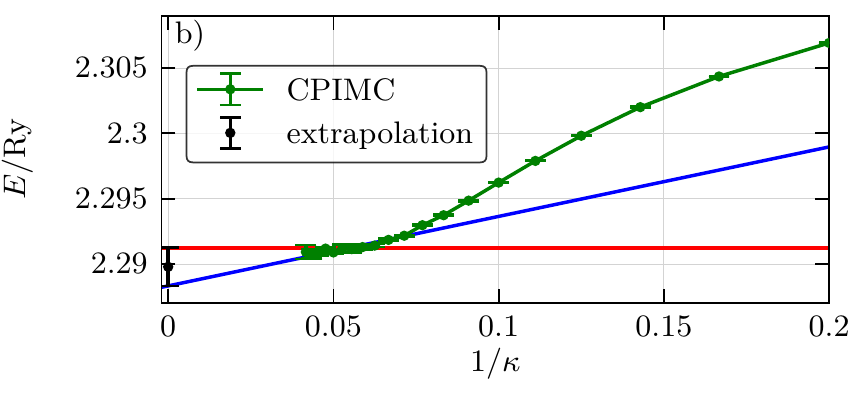}
 \caption{Convergence of the CPIMC simulations. {\bf a.} Convergence with the single-particle basis size $N_B$ for $r_s=0.4$.  The expected scaling with $N_B^{5/3}$ \cite{foulkes_n4} is well reproduced. {\bf b.} Convergence with respect to the kink potential parameter $\kappa$ (see text) and extrapolation to $1/\kappa \to 0$, corresponding to $K \to \infty$, for $r_s=1.0$ and $\theta=0.0625$. The asymptotic value is enclosed between the red and blue line.}
 \label{fig:tests}
\end{figure}

{\bf Numerical results.}
Our finite-temperature CPIMC algorithm was extensively tested for Coulomb interacting fermions in a 1D harmonic trap~\cite{schoof_cpp_11}. A first test of the present algorithm for the UEG for $N=4$ particles showed excellent agreement with exact diagonalization data~\cite{schoof_cpp14}, and the results were confirmed by density matrix QMC~\cite{foulkes_n4-comparison}. Here, we extend these simulations to $N=33$ particles. First we check the convergence with respect to the  basis size $N_B$ and show a typical case in Fig.~\ref{fig:tests}.a, for $r_s=0.4$ and four temperatures. The scaling with respect to $x=1/N_B^{-5/3}$ \cite{foulkes_n4} allows for a reliable extrapolation to $x \to 0$ and to set $N_B$ to $2109$ for all simulations, giving a relative basis incompleteness error not exceeding the statistical error ($1\sigma$ standard deviation).

With these parameters, we have performed extensive CPIMC simulations in a broad range of $r_s$- and $\Theta$-values. In contrast to DPIMC and RPIMC, our ab initio simulations (without any simplifications besides the choice of $N_B$) pose no problem for the ideal and weakly coupled UEG, up to $r_s \sim 0.4$. For larger $r_s$, we observe a rapid decrease 
of the average sign, in analogy to the harmonic oscillator case~\cite{schoof_cpp_11}. This gives rise to convergence problems of the MC algorithm in case a path with many kinks is attempted. 
We, therefore, introduce an artificial kink potential  in Eq.~(\ref{eq:z-kinks}),
%
$ V_{\kappa}(K)=[e^{-(\kappa+0.5-K)}+1]^{-1}$,
%
for calculations with $r_s > 0.4$, yielding the correct partition function in the limit $\kappa \to \infty$. Performing simulations for different $\kappa$, we generally observe a rapid convergence of the total energy allowing for an extrapolation to $1/\kappa \to 0$. 
This is demonstrated for the most difficult case ($r_s=1, \Theta=0.0625$) in Fig.~\ref{fig:tests}.b. The asymptotic value and the error estimate are computed from the two extreme cases of a horizontal and linear extrapolation. With this procedure the simulations could be extended to $r_s=1$, with the total error not exceeding $0.15\%$.

\begin{figure}
 \includegraphics{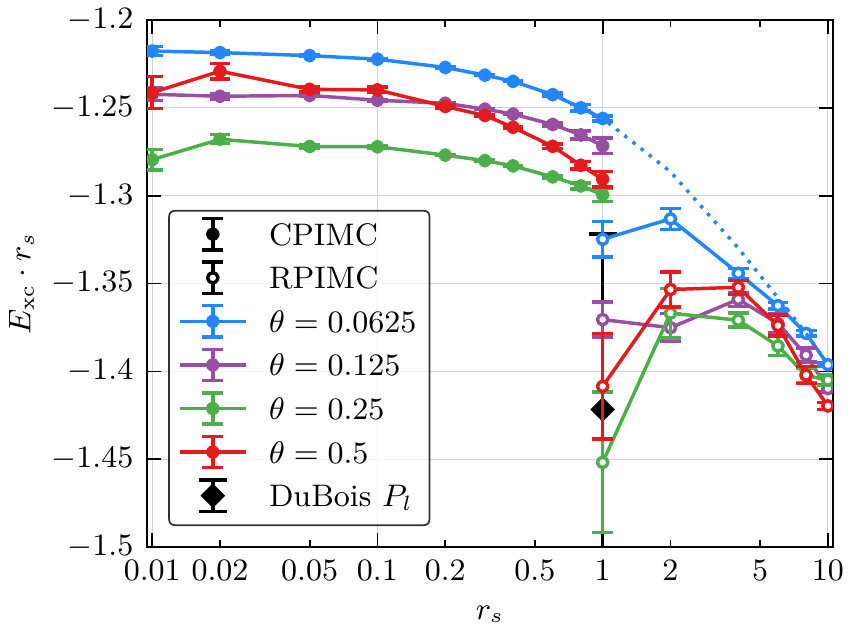}
 \caption{(Color online) Exchange-correlation energy (times $r_s$) for $33$ spin-polarized electrons and four temperatures. Comparison of our CPIMC results (full symbols with error bars \cite{fluct_small_rs}) and RPIMC results of  Ref.~\cite{brown_prl13} (open symbols).  The dotted line is an interpolation between the CPIMC and RPIMC data for $\Theta=0.0625$. Also shown is the data point of DuBois {\em et al.}~\cite{dubois14} for $\Theta=0.125$.}
 \label{fig:xc_energy_nofsc}
\end{figure}

Our simulation results for the exchange-correlation energy per electron $E_{\rm xc}$ are summarized in Fig.~\ref{fig:xc_energy_nofsc}. The data cover the whole  range $0\le r_s\le 1$ and include the ideal Fermi gas where $E_{\rm xc}r_s \to $ const (Hartree-Fock limit).
A detailed table of the various energy contributions is presented in the supplementary material~\cite{supplement}. 
A non-trivial observation is the non-monotonic temperature dependence (cf. crossing of the red and pink curves) that is in agreement RPIMC and the macroscopic fit of Ref.~\cite{karasiev_prl14}.
Interestingly, all curves seem to cross over smoothly into the  RPIMC data \cite{brown_prl13, brown-data33}, for $r_s \gtrsim 4$, as indicated by the dotted line. There is an obvious mismatch in the range $r_s \sim 1 \dots 4$. Since our curves are accurate within the given error, this discrepancy is expected to be due to the (unknown) systematic error involved in RPIMC. Also, the energy obtained within the novel approach of DuBois {\em et al.}~\cite{dubois14}, cf. the data point for $r_s=1$ and $\Theta=0.125$ in Fig.~\ref{fig:xc_energy_nofsc}, is found to be too low. 

\begin{figure}
 \includegraphics[width=85mm]{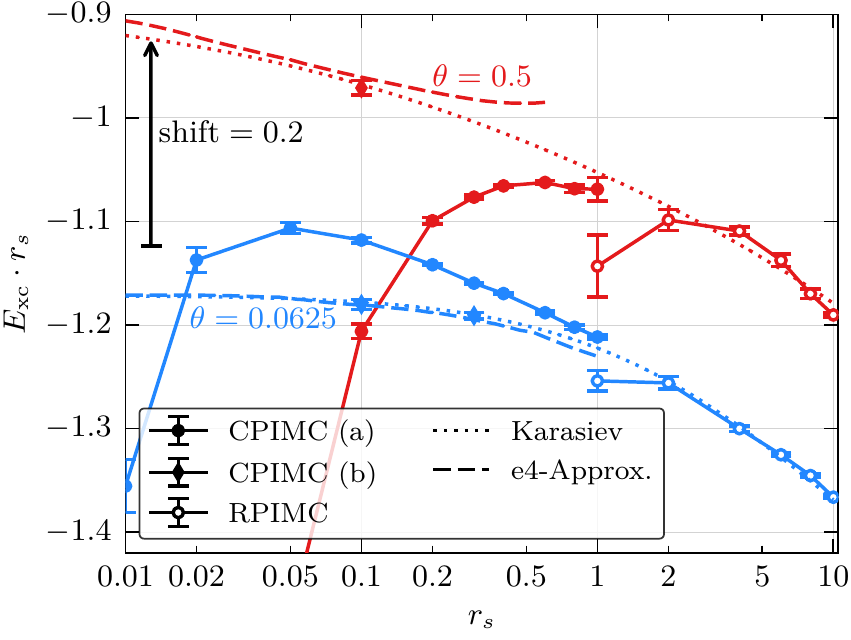}
 \caption{(Color online) Exchange correlation energy (times $r_s$) of the macroscopic polarized UEG at $\Theta=0.0625$ (blue) and $\Theta=0.5$ (red). 
Open symbols: RPIMC results~\cite{brown_prl13}. 
CPIMC (a): our results with FSC from Ref.~\cite{drummond_fsc}. RCPIMC (b): our data (3 points) with numerical extrapolation $N\to \infty$ \cite{rcpimc}. Dashes: analytical $e^4$-approximation \cite{supplement}, dots: fit of Ref.~\cite{karasiev_prl14}. For better visibility the curves for $\Theta=0.5$ are upshifted by $0.2$.}
 \label{fig:xc_energy}
\end{figure}

{\bf Macroscopic results.} Predictions for a macroscopic system, based on data for just $33$ particles, 
will inevitably lead to a loss of accuracy. Brown {\em et al.} have attempted such a mapping of their RPIMC data for $N=33$ to $N \to \infty$ and published finite size corrections (FSC) in the supplement of Ref.~\cite{brown_prl13}. A ground state formula [FSC (a)] has been presented by Drummond {\em et al.} \cite{drummond_fsc}. 
We tested both FSC, after incorporating a twist-averaging procedure in our simulations. 
For the lowest temperature, $\Theta=0.0625$ and $r_s=1$, FSC(a) leads to
 reasonable agreement with analytical approximations (see below), 
and smoothly connects to the RPIMC data, for $r_s \gtrsim 5$, cf. Fig.~\ref{fig:xc_energy}. For smaller $r_s$ and higher $\Theta$ the formula 
is not applicable. 
On the other hand, the FSC of Brown produces energies that are systematically too high and has to be discarded \cite{tw_brown_fsc}.
Due to the lack of applicable high-density FSC, we performed additional  CPIMC simulations for a broad range of particle numbers up to $N_{max}=400$, allowing for an extrapolation to the macroscopic limit, for $r_s=0.1$ and $\Theta=0.0625;\;\Theta=0.5$. For $\Theta=0.0625$ an additional point at $r_s=0.3$  could be obtained \cite{rcpimc}, cf. the diamonds in Fig.~\ref{fig:xc_energy}. These accurate data will be a suitable starting point for the construction of FSC formulas that are applicable at high densities.

To obtain independent analytical results for the macroscopic UEG, we now compute the exchange correlation energy including, in addition to Hartree-Fock~\cite{karasiev_prl14}, the two second order diagrams (Montroll Ward and $e^4$)~\cite{supplement}. 
The two results [cf. Fig.~\ref{fig:xc_energy}], converge for low $r_s$, eventually reaching the Hartree-Fock asymptote (horizontal line). 
For $r_s \gtrsim 0.1$,  MW and $e^4$ start to deviate from one another, 
and we expect the exact result to be enclosed between the two~\cite{supplement}. Reliable predictions are possible up to $r_s \sim 0.8$, for $\Theta = 0.0625$, and $r_s \sim 0.55$, for $\Theta = 0.5$
\cite{supplement}.
In Figure~\ref{fig:xc_energy} we also include the fit of Ref.~\cite{karasiev_prl14}
that shows, overall, a very good behavior, but is too low at $r_s \to 0$, with the deviations growing with $\Theta$~\cite{supplement}. 
%


{\em To summarize}, we have presented first-principle configuration PIMC results for the UEG at finite temperature that have no sign problem at high degeneracy, $r_s \lesssim 0.4$, and allow for reliable predictions up to $r_s=1$. 
This makes CPIMC a perfect complementary approach to direct fermionic PIMC and to RPIMC that cannot access high densities, and our results indicate that the previous RPIMC data are not reliable in the range $r_s \lesssim 4$. The present results will be important for dense quantum plasmas at finite temperatures, that are relevant for warm dense matter, in general, and for ICF, in particular. 
Since here the electrons are typically unpolarized we tested our CPIMC approach for this case. Although the sign problem is more severe than for the polarized situation CPIMC is well capable to produce very accurate ab initio finite temperature results that smoothly connect to available $T=0$ data~\cite{supplement}.
A systematic analysis is presently under way. 
%

Finally, the obtained accurate results for the exchange-correlation energy provide benchmarks for finite temperature DFT, RPIMC \cite{brown_prl13}, novel independent QMC simulations \cite{foulkes_n4,dubois14} and  analytical fits~\cite{karasiev_prl14}. Although the fermion sign problem is not removed by our approach, the proposed combination of CPIMC with DPIMC (or RPIMC) provides, for the UEG, a practical way around this problem.

We acknowledge stimulating discussions with V.~Filinov, J.W.~Dufty,  V.V.~Karasiev and S.~Trickey and E. Brown for providing information on the FSC (a) used in Ref.~\cite{brown_prl13}.
This work is supported by the Deutsche Forschungsgemeinschaft via grant BO1366/10 and by grant SHP006 for supercomputing time at the HLRN.


\begin{thebibliography}{26}
\bibitem{relativistic} In this paper we restrict ourselves to the non-relativistic jellium model.
\bibitem{ctmc-comment} This concept is also used in zero temperature full configuration Monte Carlo \cite{alawi_jcp12, alawi_prb12}.
\bibitem{kink_note}
Due to momentum conservation, only two-particle excitations $s_i$ are possible 
(for the general case, see Ref.~\cite{simon_springer14}), and the restriction to closed paths requires $K\ne 1$.
\bibitem{fluct_small_rs} The fluctuations of $E_{xc}$ at small $r_s$ arise from the vanishing of $E_{xc}$, compared to the total energy, for $r_s \to 0$. The latter is always well converged.
\bibitem{rcpimc} For $r_s=0.3$ the extrapolation to $N=150$ was achieved with an additional approximation that includes only even kink numbers. It is found to be accurate for low temperatures.
\end{thebibliography}
\end{document}